\documentclass[twocolumn,10pt,twoside]{IEEEtran}
\pdfoutput=1
\usepackage[T1]{fontenc}
\usepackage{flushend}
\usepackage{grffile}
\usepackage{bm}
\usepackage{amsmath}
\usepackage{amsthm}
\usepackage{amssymb}
\usepackage{graphicx}
\PassOptionsToPackage{normalem}{ulem}
\usepackage{ulem}
\usepackage[unicode=true,
 bookmarks=true,bookmarksnumbered=true,bookmarksopen=true,bookmarksopenlevel=1,
 breaklinks=false,pdfborder={0 0 0},backref=false,colorlinks=false,hidelinks]
 {hyperref}
\hypersetup{pdftitle={Recovery of non-linear cause-effect relationships from linearly mixed neuroimaging data},
 pdfpagelayout=OneColumn, pdfnewwindow=true, pdfstartview=XYZ, plainpages=false,
 pdfauthor={Sebastian Weichwald, Arthur Gretton, Bernhard Schölkopf, Moritz Grosse-Wentrup}}

\theoremstyle{definition}
\newtheorem{defn}{\protect\definitionname}
\providecommand{\definitionname}{Definition}

\usepackage{setspace}
\usepackage{pgfplots}
\tikzset{node/.style={circle, draw,font=\sffamily\bfseries, minimum size=2.5em, text centered, line width=1pt}}
\tikzset{nodeh/.style={circle, dashed, draw,font=\sffamily\bfseries, minimum size=2.5em, text centered, line width=1pt}}
\tikzset{d/.style={line width=1pt}}

\usepackage{centernot}
\def\independenT#1#2{\mathrel{\rlap{$#1#2$}\mkern2mu{#1#2}}}
\newcommand{\indep}{\protect\mathpalette{\protect\independenT}{\perp}}
\newcommand{\dep}{\centernot\indep}

\newcommand\copyrighttext{
This is the author's version of an article that is published in \textit{Pattern Recognition in Neuroimaging (PRNI), International Workshop on,} 1--4, 2016, \href{http://dx.doi.org/10.1109/PRNI.2016.7552331}{doi: 10.1109/PRNI.2016.7552331}.\\
Copyright (c) 2016 IEEE. Personal use is permitted. For any other purposes, permission must be obtained from the IEEE by emailing pubs-permissions@ieee.org.
}

\newcommand\copyrightnotice{%
\begin{tikzpicture}[remember picture,overlay]
\node[anchor=south] at (current page.south) {{\parbox{\textwidth}{\centering\scriptsize\copyrighttext}}};
\end{tikzpicture}%
}

\begin{document}

\title{Recovery of non-linear cause-effect relationships from linearly mixed
neuroimaging data}

\author{Sebastian Weichwald, Arthur Gretton, Bernhard Schölkopf, Moritz Grosse-Wentrup\thanks{SW, BS, and MGW are with the Empirical Inference Department,
Max Planck Institute for Intelligent Systems, Tübingen, Germany, e-mail:
\protect\href{}{[sweichwald, bs, moritzgw]@tue.mpg.de}.}\thanks{AG is with the Gatsby Computational Neuroscience Unit,
Sainsbury Wellcome Centre, London, United Kingdom, e-mail: \protect\href{mailto:arthur.gretton@gmail.com}{arthur.gretton@gmail.com}.}}

\maketitle
\copyrightnotice

\begin{abstract}
Causal inference concerns the identification of cause-effect relationships
between variables. However, often only linear combinations of variables
constitute meaningful \emph{causal} variables. For example, recovering
the signal of a cortical source from electroencephalography requires
a well-tuned combination of signals recorded at multiple electrodes.
We recently introduced the MERLiN (Mixture Effect Recovery in Linear Networks)
algorithm that is able to recover, from an observed linear mixture,
a causal variable that is a \emph{linear} effect of another given
variable. Here we relax the assumption of this cause-effect relationship
being linear and present an extended algorithm that can pick up \emph{non-linear}
cause-effect relationships. Thus, the
main contribution is an algorithm (and ready to use code) that has
broader applicability and allows for a richer model class.
Furthermore, a comparative analysis indicates that the assumption of linear cause-effect
relationships is not restrictive in analysing electroencephalographic
data.
\end{abstract}


\section{Introduction}

Causal inference requires causal variables. However, not always do
the variables in a dataset specify the candidate causal relata. In
electroencephalography (EEG) studies, for example, what is measured
at electrodes placed on the scalp is instantaneously and
linearly superimposed electromagnetic activity of sources in the brain
\cite{Nunez2006}. Standard causal inference methods require to first
recover the cortical sources from the observed electrode signals \cite{grosse2009understanding}. This is disadvantageous. First, any source localisation procedure is prone to modelling errors, which may distort the true cause-effect relationships between cortical sources. Second, source localisation enlarges the data dimensionality by roughly two orders of magnitude, which leads to increased computational complexity.

We recently proposed a novel idea to construct causal variables, i.\,e.,
recover cortical sources, by directly optimising for statistical in-
and dependences that imply a certain cause-effect relationship \cite{weichwald2015merlin}. The
linear MERLiN algorithm can -- skipping potentially error prone modelling
steps -- establish a \emph{linear} cause-effect relationship between
brain state features that are observed only as part of a linear mixture.
This allows for computationally efficient insights into brain networks beyond those readily
obtained from encoding and decoding models trained on pre-defined
variables \cite{Weichwald2015}. The linear MERLiN algorithm, however, is unable to reconstruct cortical sources with non-linear cause-effect relationships.

Here we present the \emph{non-linear }MERLiN algorithm and relax the
assumption of linear cause-effect relationships. By integrating kernel ridge regression and a non-linear independence test, the extended algorithm can capture any higher order dependence. We compare the results of our linear- and non-linear MERLiN algorithms on EEG data for which cause-effect relationships have previously only been computed by an exhaustive search approach \cite{grosse2015identification} and find no qualitative differences. The contribution of this work is thus two-fold. First, we provide an algorithm to learn non-linear cause-effect relationships from linear mixtures of causal variables, and, second, we provide empirical evidence that linear methods suffice to identify cause-effect relationships within individual EEG frequency bands. The Python implementation is available at \href{https://github.com/sweichwald/MERLiN}{https://github.com/sweichwald/MERLiN}.

\section{Methods}

\subsection{Causal Bayesian Networks\label{sub:CBNs}}

We briefly introduce the main aspects of Causal Bayesian Networks
(CBNs). For an exhaustive treatment see \cite{Spirtes2000,Pearl2009}.
The important advantage of this framework over methods based on information
flow is that it yields testable predictions on the impact of interventions
\cite{eichler2010granger,lizier2010differentiating}.
\begin{defn}[Structural Equation Model]
We define a \emph{structural equation model (SEM) $\mathcal{{S}}$}
as a set of equations $X_{i}=f_{i}(\mathbf{{PA}}_{i},N_{i}),\ i\in\mathbb{{N}}_{1:s}$
where the so-called noise variables are independently distributed
according to $\mathbb{{P}}^{N_{1},...,N_{s}}=\mathbb{{P}}^{N_{1}}\cdots\mathbb{{P}}^{N_{s}}$.
For $i\in\mathbb{{N}}_{1:s}$ the set $\mathbf{{PA}}_{i}\subseteq\{X_{1},...,X_{s}\}\setminus X_{i}$
contains the so-called parents of $X_{i}$ and $f_{i}$ describes
how $X_{i}$ relates to the random variables in $\mathbf{{PA}}_{i}$
and $N_{i}$. The induced joint distribution is denoted by $\mathbb{{P}}^{\mathcal{{S}}}\triangleq\mathbb{{P}}^{X_{1},...,X_{s}}$.

Replacing at least one of the functions $f_{i},\ i\in\mathbb{{N}}_{1:s}$
by a constant $\spadesuit$ yields a new SEM. We say $X_{i}$ has
been intervened on, which is denoted by $\operatorname{do}(X_{i}=\spadesuit)$,
leads to the SEM $\mathcal{{S}}|\operatorname{do}(X_{i}=\spadesuit)$,
and induces the \emph{interventional distribution} $\mathbb{{P}}^{\mathcal{{S}}|\operatorname{do}(X_{i}=\spadesuit)}\triangleq\mathbb{{P}}^{X_{1},...,X_{s}|\operatorname{do}(X_{i}=\spadesuit)}$.
\end{defn}

\begin{defn}[Cause and Effect]
\label{def:causeeffect}$X_{i}$ is a \emph{cause} of $X_{j}$ ($i,j\in\mathbb{{N}}_{1:s},\ i\neq j$)
wrt. a SEM $\mathcal{{S}}$ iff there exists $\heartsuit\in\mathbb{{R}}$
such that $\mathbb{{P}}^{X_{j}|\operatorname{do}(X_{i}=\heartsuit)}\neq\mathbb{{P}}^{X_{j}}$.\footnote{$\mathbb{{P}}^{X_{j}|\operatorname{do}(X_{i}=\heartsuit)}$ and $\mathbb{{P}}^{X_{j}}$
denote the marginal distributions of $X_{j}$ corresponding to $\mathbb{{P}}^{\mathcal{{S}}|\operatorname{do}(X_{i}=\heartsuit)}$
and $\mathbb{{P}}^{\mathcal{{S}}}$ respectively.} $X_{j}$ is an \emph{effect} of $X_{i}$ iff $X_{i}$ is a cause
of $X_{j}$. Often the considered SEM $\mathcal{{S}}$ is omitted
if it is clear from the context.
\end{defn}
For each SEM $\mathcal{{S}}$ there is a corresponding graph $\mathcal{{G_{S}}}(V,E)$
with $V\triangleq\{X_{1},...,X_{s}\}$ and $E\triangleq\{(X_{i},X_{j}):\ X_{i}\in\mathbf{{PA}}_{j},\ X_{j}\in V\}$
that has the random variables as nodes and directed edges pointing
from parents to children. We employ the common assumption that this
graph is acyclic, i.e., $\mathcal{{G_{S}}}$ will always be a directed
acyclic graph (DAG).

So far a DAG $\mathcal{{G_{S}}}$ simply depicts all parent-child
relationships defined by the SEM $\mathcal{{S}}$. Missing directed
paths indicate missing cause-effect relationships. In order to specify
the link between statistical independence (denoted by $\indep$) wrt.
the joint distribution $\mathbb{{P}}^{\mathcal{{S}}}$ and properties
of the DAG $\mathcal{G_{S}}$ (representing a SEM $\mathcal{S}$)
we need the following definition.
\begin{defn}[d-separation]
For a fixed graph $\mathcal{{G}}$ disjoint sets of nodes $A$ and
$B$ are \emph{d-separated} by a third disjoint set $C$ (denoted
by $A\perp_{\text{{d-sep}}}B|C$) iff all pairs of nodes $a\in A$
and $b\in B$ are d-separated by $C$. A pair of nodes $a\neq b$
is d-separated by $C$ iff every path between $a$ and $b$ is blocked
by $C$. A path between nodes $a$ and $b$ is blocked by $C$ iff
there is an intermediate node $z$ on the path such that (i) $z\in C$
and $z$ is tail-to-tail ($\gets z\to$) or head-to-tail ($\to z\to$),
or (ii) $z$ is head-to-head ($\to z\gets$) and neither $z$ nor
any of its descendants is in $C$.

Conveniently, assuming faithfulness\footnote{Intuitively, this is saying that conditional independences are due
to the causal structure and not accidents of parameter values \cite[p. 9]{Spirtes2000};
more formally the assumption reads $A\perp_{\text{{d-sep}}}B|C\impliedby A\indep B|C$.} (and exploiting Markovianity\footnote{The distribution $\mathbb{P}^{\mathcal{S}}$ generated by a SEM $\mathcal{S}$
is Markov wrt. $\mathcal{G_{\mathcal{S}}}$ (cf. \cite[Theorem 1.4.1]{Pearl2009}
for a proof), i.\,e., $A\perp_{\text{{d-sep}}}B|C\implies A\indep B|C$.}) we have the following one-to-one correspondence between d-separation
and conditional independence statements:
\begin{align*}
A\perp_{\text{{d-sep}}}B|C\quad & \iff\quad A\indep B|C
\end{align*}
Summing up, we have defined interventional causation in terms of SEMs
and have seen how a SEM gives rise to a DAG. This DAG has two convenient
features. Firstly, the DAG yields a visualisation that allows to easily
grasp missing cause-effect relationships that correspond to missing
directed paths. Secondly, since we assume faithfulness, d-separation
properties of this DAG are equivalent to conditional independence
properties of the joint distribution. Thus, conditional independences
translate into causal statements, e.g. `a variable becomes independent
of all its non-effects given its immediate causes' or `cause and effect
are marginally dependent'. Furthermore, the causal graph $\mathcal{G_{S}}$
can be identified from conditional independences observed in $\mathbb{P}^{\mathcal{S}}$
-- at least up to a so-called \emph{Markov equivalence class,} the
set of graphs that entail the same conditional independences \cite{Verma90}.
\end{defn}

\subsection{Formal problem description}

In the following, the variables $S$, $C_{1},...,C_{d}$, and $F_{1},...,F_{d}$
may be thought of as a stimulus variable, the activity of multiple
cortical sources, and the EEG channel recordings respectively. We
aim at recovering an effect of a pre-defined target variable $C_{1}=\bm{v}^{\top}F$.
The terminology introduced in Section \ref{sub:CBNs} allows to precisely
state the problem as follows.

\subsubsection{Assumptions\label{sub:assumptions}}

\begin{figure}[tbh]
\begin{center}
\-\newline
\begin{tikzpicture}[scale=.7, every node/.style={scale=.7}]
    \node[node] (s) at(-0.5,0) {$S$};
    \node[node] (c1) at(1.5,0) {$C_1$};
    \node[node] (c2) at(3,0) {$C_2$};
    \node[node] (c3) at(1.5,-1.5) {$C_3$};
    \node[nodeh] (h) at(3,-1.5) {$h_1$};
    \node[node] (c4) at(4.5,-1.5) {$C_4$};
    \node[node] (c5) at(6,0) {$C_5$};
    \node at(6,-.65) {$\vdots$};
    \node[node] (cd) at(6,-1.5) {$C_d$};

    \draw[->,d] (s) -- (c1);
    \draw[->,d] (s) -- (c3);
    \draw[->,d] (c1) -- (c2);
    \draw[->,d] (h) -- (c1);
    \draw[->,d] (h) -- (c4);
\end{tikzpicture}\end{center}\caption{Example graph where $h_{1}$ is a hidden variable.}
\label{fig:example}
\end{figure}
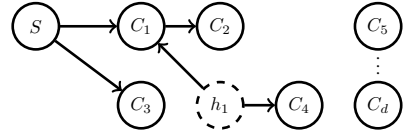

Let $S$ and $C_{1},...,C_{d}$ denote (finitely many) random variables.
We assume existence of a SEM $\mathcal{{S}}$, potentially with additional
unobserved variables $h_{1},...,h_{l}$, that induces $\mathbb{{P}}^{\mathcal{{S}}}=\mathbb{{P}}^{S,C_{1},...,C_{d},h_{1},...,h_{l}}$.
We refer to the corresponding graph $\mathcal{G_{S}}$ as the \emph{true}
\emph{causal graph} and call its nodes \emph{causal variables}. We
further assume that
\begin{itemize}
\item $S$ affects $C_{2}$ indirectly via $C_{1}$,\footnote{By saying a variable $X$ causes $Z$ indirectly via $Y$ we imply
(a) existence of a path $X\dasharrow Y\dasharrow Z$, and (b) that
there is no path $X\dasharrow Z$ without $Y$ on it (this also excludes
the edge $X\to Z$).}
\item there are no edges in $\mathcal{{G_{S}}}$ pointing into $S$.\footnote{This condition can for example be ensured by randomising $S$.}
\end{itemize}
Importantly, we do not require that the structural equation that relates
$C_{2}$ to its parents is \emph{linear }in $C_{1}$. Figure \ref{fig:example}
depicts an example of how $\mathcal{{G_{S}}}$ might look like.

\subsubsection{Given data}
\begin{itemize}
\item $\bm{v}\in\mathbb{{R}}^{d}$ such that $C_{1}=\bm{v}^{\top}F$
\item $m$\emph{ iid}\footnote{independent and identically distributed}\emph{
}samples $\bm{S}=[s_{1},...,s_{m}]^{\top}$ of $S$ and $\bm{F}=[f_{i,j}]_{i=1:m,j=1:d}$
of $F$ where $F\triangleq[F_{1},...,F_{d}]^{\top}=\bm{A}C$ is the
observed linear mixture of the causal variables $C\triangleq[C_{1},...,C_{d}]^{\top}$
and $\bm{A}\in\mathbb{{R}}^{d\times d}$ the mixing matrix
\end{itemize}

\subsubsection{Desired output}

Find $\bm{w}\in\mathcal{B}^{d}\triangleq\{\bm{x}\in\mathbb{R}^{d}:||\bm{x}||=1\}$
such that $aC_{i}=\bm{w}^{\top}F$ where $C_{i}$ is an effect of
$C_{1}$ ($i\in\mathbb{{N}}_{2:d},a\in\mathbb{R}\setminus\{0\}$).
For the graph shown in Figure~\ref{fig:example} recovery of the
causal variable $C_{2}$ is a valid solution.

\subsection{Strategy\label{sec:idea}}

Our approach leverages the following causal inference rule that --
under the assumptions in Section \ref{sub:assumptions} -- applies
to a \emph{causal} variable $C_i$ (cf. \cite{grosse2015identification}).

\smallskip{}

\paragraph*{Causal Inference Rule}

If $C_i\indep S|C_{1}$ and $Y\dep C_{1}$, then $S$ indirectly affects
$C_i$ via $C_{1}$. In particular, a causal path $C_{1}\dashrightarrow C_i$
exists.

\smallskip{}

The idea is to recover the sought-after variable from the mixture
$F$ by optimising for these statistical properties. Thence, the general
strategy relies on solving an optimisation problem of the form\footnote{To attenuate the signal of $C_{1}$ we restrict search onto the orthogonal
complement $\bm{v}_{\perp}\cap\mathcal{B}^{d}=\{\bm{x}\in\mathbb{R}^{d}:||\bm{x}||=1,\bm{x}\perp\bm{v}\}$
which is diffeomorphic to $\mathcal{B}^{d-1}$.}
\[
\max_{\bm{w}\in\mathcal{B}^{d-1}}\operatorname{dep}(C_{1},Y_{\bm{w}})-\operatorname{dep}(S,Y_{\bm{w}}|C_{1})
\]
where $Y_{\bm{w}}=\bm{w}^{\top}F$ and $\operatorname{dep}$ denotes
a (conditional) dependence criterion that estimates from empirical
samples the strength of association between the two variables.

\subsection{Non-Linear MERLiN algorithm}

The linear MERLiN algorithm uses the partial correlations $\rho_{C_{1},Y_{\bm{w}}|S}$
and $\rho_{S,Y_{\bm{w}}|C_{1}}$ for the terms $\operatorname{dep}(C_{1},Y_{\bm{w}})$
and $\operatorname{dep}(S,Y_{\bm{w}}|C_{1})$ in the objective function.
As such only \emph{linear} dependence between $C_{1}$ and $Y_{\bm{w}}$
can be detected while remaining higher-order dependences between $S$
and $Y_{\bm{w}}$ given $C_{1}$ may go undetected. A general kernel-based
independence criterion, the Hilbert-Schmidt Independence Criterion
(HSIC)~\cite{Gretton2008}, and a regression-based conditional independence
criterion (cf. \cite{grosse2015identification}) in conjunction with
kernel ridge regression~\cite{saunders1998ridge} allow extension
to \emph{non-linear} dependences.

\smallskip{}

\paragraph*{Regression-Based Conditional Independence Criterion}

If there exists a (regression) function $r$ such that $Y_{\bm{w}}-r(C_{1})\indep(S,C_{1})$
then $S\indep Y_{\bm{w}}|C_{1}$.

\smallskip{}

The non-linear MERLiN algorithm solves the following optimisation
problem 
\[
\max_{(\bm{w},\sigma,\theta)\in\mathcal{B}^{d-1}\times\mathbb{R}\times\mathbb{R}}\operatorname{HSIC}(C_{1},Y_{\bm{w}})-\operatorname{HSIC}\left((S,C_{1}),R_{\bm{w},\sigma,\theta}\right)
\]
where $\operatorname{HSIC}(A,B)$ denotes the empirical HSIC estimate\footnote{We compute the empirical HSIC estimate based on the Gaussian kernel
$k(x,y)=\exp(-\delta^{-1}||x-y||^{2})$ where the kernel size $\delta$
is determined by the median distance between points in input space
\cite{Gretton2008}.} and $R_{\bm{w},\sigma,\theta}$ corresponds to the residuals $Y_{\bm{w}}-r(C_{1})$
using kernel ridge regression with Gaussian kernel of width $|\sigma|$
and ridge regression parameter $|\theta|$. To temper overfitting,
the sample is split into three partitions; the residuals of the $i^{\text{th}}$
partition are obtained by using the kernel ridge regression function
obtained on the remaining partitions. The regression parameters $\sigma$
and $\theta$ are also being optimised over to allow an optimal regression
fit wrt.\ witnessing conditional independence and hence minimising
the second summand in the objective function.

Implementing the objective function in Theano \cite{Bergstra2010,Bastien2012},
we use the Python toolbox Pymanopt \cite{townsend2016} to run optimisation
on the product manifold $\mathcal{B}^{d-1}\times\mathbb{R}\times\mathbb{R}$
using a steepest descent algorithm with standard back-tracking line-search.
This approach is exact and efficient, relying on automated differentiation
and respecting the manifold geometry.

\subsection{Application to EEG data}

We consider EEG trial-data of the form $\widetilde{\bm{F}}\in\mathbb{R}^{d\times m\times n}$
where $d$ denotes the number of electrodes, $m$ the number of trials,
and $n$ the length of the time series $\widetilde{\bm{F}}_{i,j,1:n}$
for each electrode $i\in\mathbb{N}_{1:d}$ and each sample $j\in\mathbb{N}_{1:m}$;
that is $\widetilde{\bm{F}}$ holds $m$ iid samples of a $\mathbb{R}^{d\times n}$-valued random variable~$\widetilde{F}$.
Analyses of EEG data commonly focus on trial-averaged log-bandpower
in a particular frequency band. Accordingly, applying our algorithms
to EEG data we aim to identify a linear combination $\bm{w}\in\mathcal{B}^{d}$
such that the log-bandpower of the resulting one-dimensional trial
signals $\bm{w}^{\top}\widetilde{F}$ is a causal
effect of the log-bandpower of the one-dimensional trial signals~$\bm{v}^{\top}\widetilde{F}$.

However, the two operations of computing the log-bandpower and taking
a linear combination do not commute. The log-bandpower computation
needs to be switched into the objective function described above.
This is accomplished by letting $Y_{\bm{w}}=\operatorname{logbp}(\bm{w}^{\top}\widetilde{F})$
and $C_{1}=\operatorname{logbp}(\bm{v}^{\top}\widetilde{F})$
where $\operatorname{logbp}$ denotes the operation of applying a
Hanning window and computing the average log-bandpower in a specified
frequency range for the given time series.

\section{Comparative EEG analysis}

\subsection{Experimental data}

We applied the linear MERLiN algorithm and its non-linear extension
to EEG data recorded during a neurofeedback experiment \cite{GrosseWentrup2014}.
In this study the $\gamma$-log-bandpower ($55$--$85$
Hz) in the right superior parietal cortex (SPC) was provided as feedback
signal and subjects were instructed to up- or down-regulate the bandpower.
$3$ subjects were recorded in $2$ sessions each and each session had $60$ trials á $60$ seconds.

The data of one session consists of a stimulus vector $\bm{S}\in\{-1,+1\}^{60\times1}$,
a spatial filter $\bm{v}\in\mathbb{R}^{121\times1}$ that was used
to extract the feedback signal, and a tensor $\widetilde{\bm{F}}\in\mathbb{R}^{121\times60\times15000}$
that holds the time series (of length $15000$) for each channel and
trial. The reader is referred to \cite{GrosseWentrup2014} for more
details on the experimental data.

\subsection{How to compare to previous results}

We compare our MERLiN algorithms against a causal analysis of this
neurofeedback experiment that is based on source localisation in combination with an exhaustive search procedure~\cite{grosse2015identification}. The hypothesis
was, based on transcranial magnetic stimulation studies \cite{Chen2013},
that $\gamma$-oscillations in the SPC modulate $\gamma$-oscillations
in the medial prefrontal cortex (MPC). We briefly
describe this exhaustive search approach.

First, the signal of $K\triangleq15028$ dipoles across
the cortical surface was extracted using a LCMV beamformer and a three-shell spherical
head model \cite{Mosher1999}. Then, the authors applied their newly
introduced stimulus-based causal inference (SCI) algorithm to assess
for every dipole whether its $\gamma$-log-bandpower is a \emph{linear
}causal effect of the $\gamma$-log-bandpower in the SPC. Group
results were summarised in a vector $\bm{g}_{\text{SCI}}\in\mathbb{R}^{K\times1}$
where the $i^{\text{th}}$ entry denotes the percentage of dipoles
within a certain radius that were found to be modulated by the SPC.
The results of this exhaustive search analysis, visualising $\bm{g}_{\text{SCI}}$
on hemisphere plots, supported the hypothesis that the MPC is a \emph{linear}
causal effect of the SPC. The reader is referred to \cite{grosse2015identification}
for more details.

In contrast to exhaustive search, both our linear MERLiN algorithm
as well as its non-linear extension aim at immediately recovering
the causal effect by optimising a linear combination $\bm{w}$ of
electrode signals\footnote{Since there were only $60$ samples per session we decided to select
a subset of $33$ EEG channels distributed across the scalp
(according to the $10\text{--}20$ system). Hence, for each recording
session we obtained a spatial filter $\bm{w}\in\mathbb{R}^{33\times1}$.}. To allow for a qualitative comparison of our results with the results summarised
by the vector $\bm{g}_{\text{SCI}}$ we derive for each $\bm{w}$
a vector $\bm{g}\in\mathbb{R}^{K\times1}$. This vector represents
the involvement of each cortical dipole in the recovered signal and
is derived from $\bm{w}$ as follows. First, a scalp topography is
obtained via $\bm{a}\propto\Sigma\bm{w}$ where the $i^{\text{th}}$
entry of $\Sigma\bm{w}$ is the covariance between the $i^{\text{th}}$
EEG channel and the source that is recovered by $\bm{w}$ \cite[Equation (7)]{Haufe2014}.
Here $\Sigma$ denotes the session-specific covariance matrix in the
$\gamma$-frequency band. Second, the dipole involvement vector $\bm{g}$
is obtained from $\bm{a}$ via dynamic statistical parametric mapping
(dSPM; with identity noise covariance matrix) \cite{Dale2000}. Group
results are obtained as average of the individual dipole involvement
vectors.

\subsection{Experimental results}

\begin{figure}[tbh]
\begin{minipage}[t]{0.49\linewidth}%
\begin{flushright}
{\footnotesize{}Left hemisphere\quad{}}
\par\end{flushright}{\footnotesize \par}

\begin{flushright}
\rotatebox{90}{\footnotesize\quad Lateral view} \includegraphics[width=0.7\linewidth]{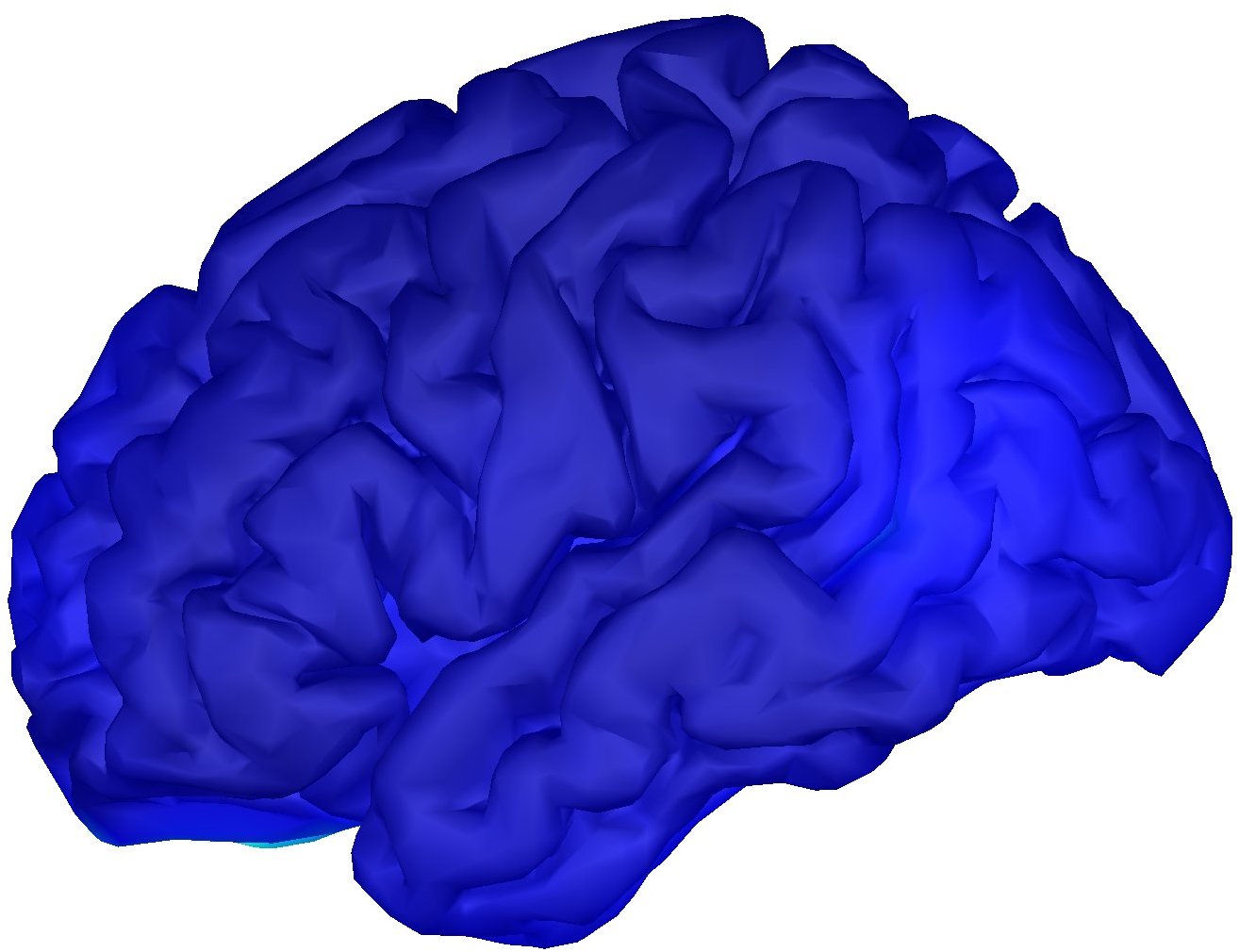}
\par\end{flushright}%
\end{minipage}%
\begin{minipage}[t]{0.49\linewidth}%

\begin{flushleft}
{\footnotesize{}\quad{}Right hemisphere}
\par\end{flushleft}{\footnotesize \par}

\begin{flushleft}
\includegraphics[width=0.7\linewidth]{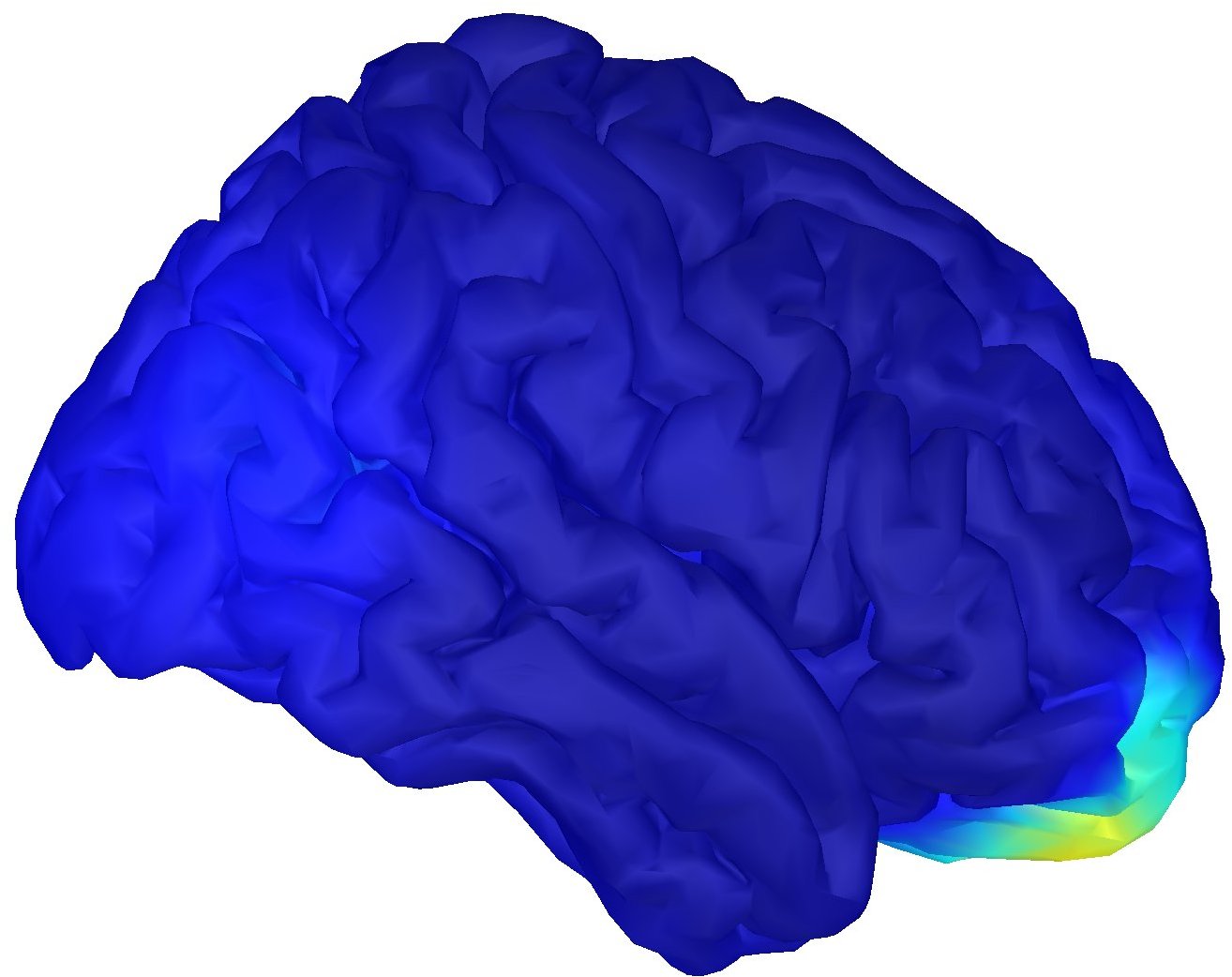}
\par\end{flushleft}%
\end{minipage}

\begin{minipage}[t]{0.49\linewidth}%
\begin{flushright}
\rotatebox{90}{\footnotesize\quad Medial view} \includegraphics[width=0.7\linewidth]{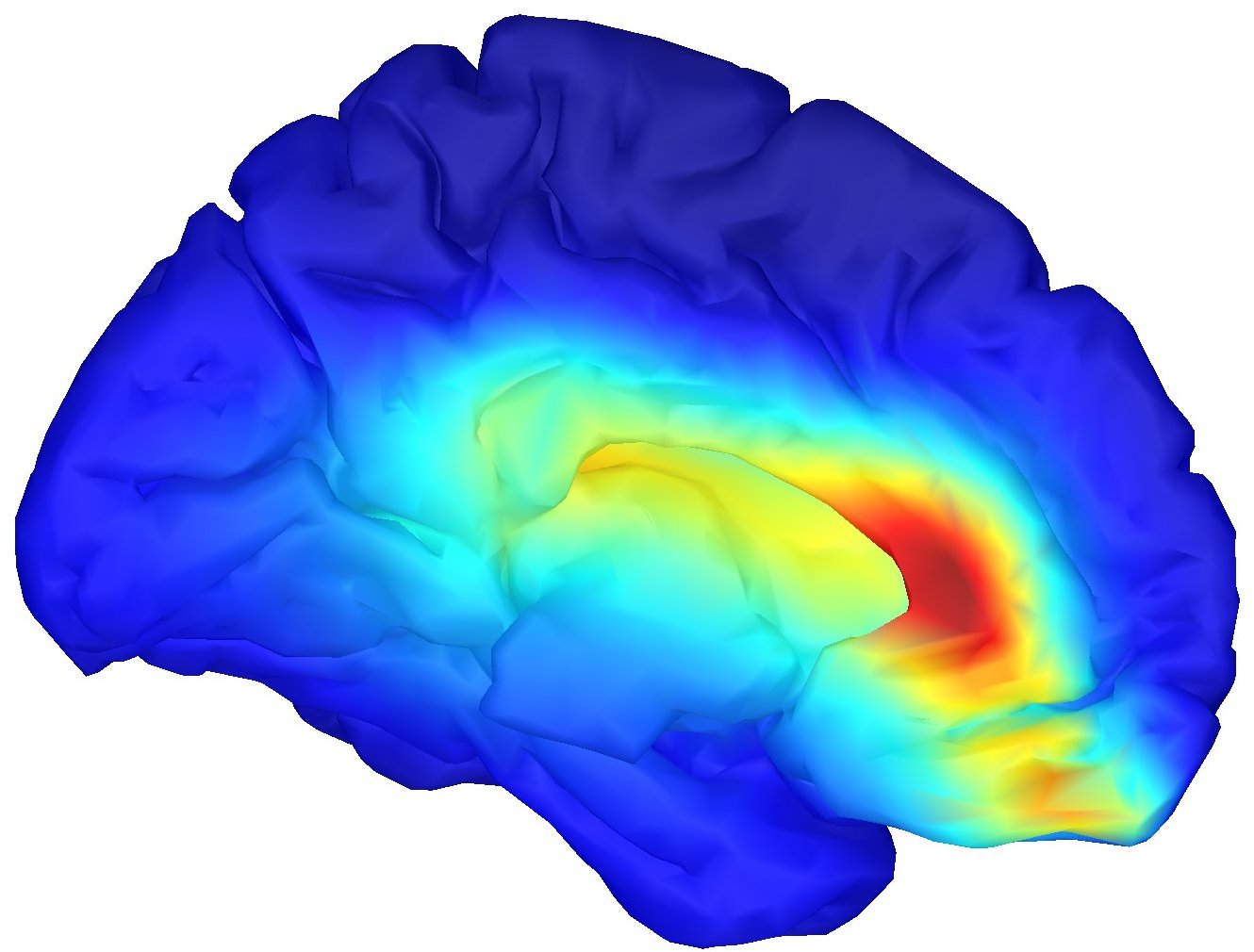}
\par\end{flushright}%
\end{minipage}%
\begin{minipage}[t]{0.49\linewidth}%
\begin{flushleft}
\includegraphics[width=0.7\linewidth]{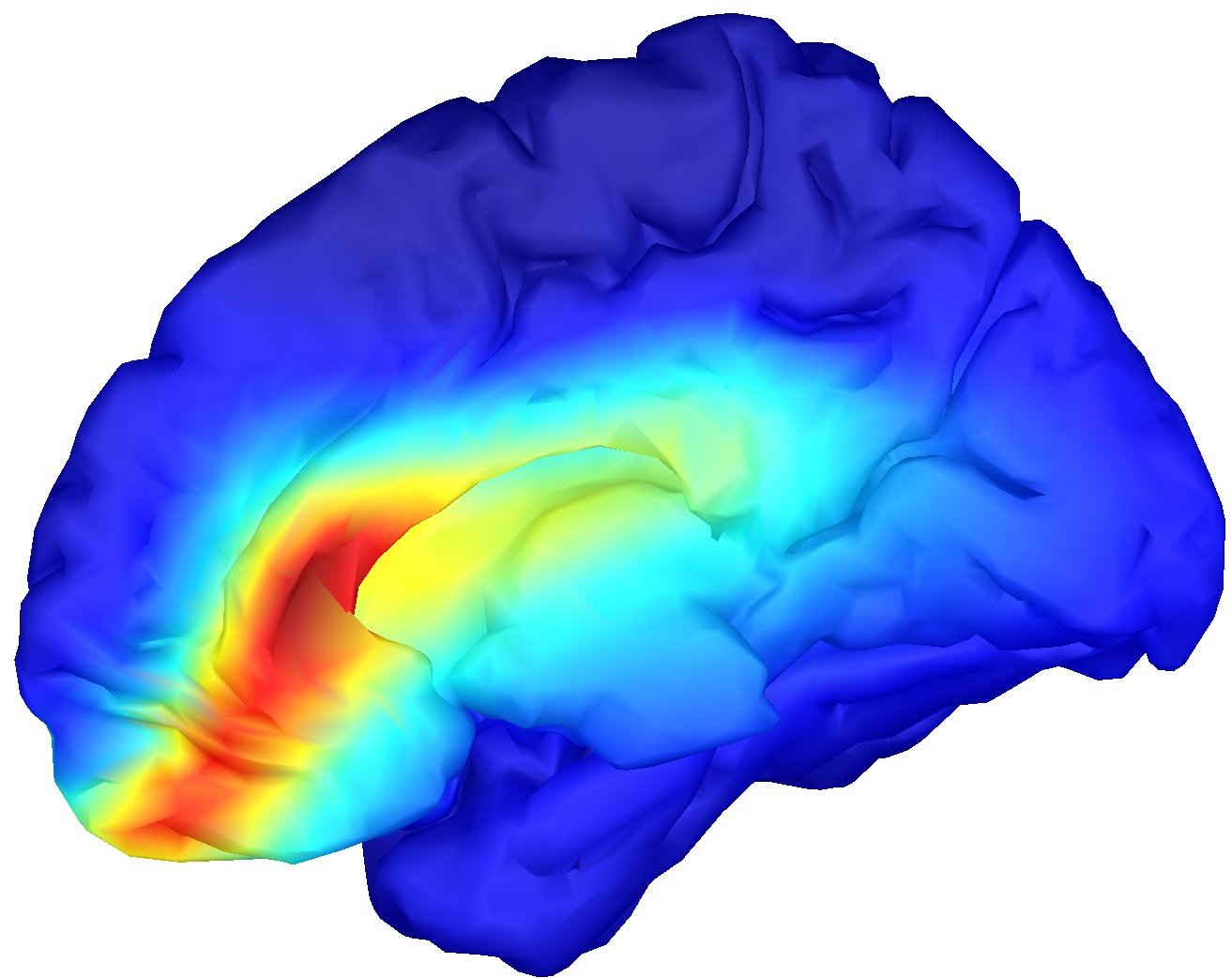}
\par\end{flushleft}%
\end{minipage}
\centerline{(a) non-linear}

\vspace{1em}

\begin{minipage}[t]{0.49\linewidth}%
\begin{flushright}
{\footnotesize{}Left hemisphere\quad{}}
\par\end{flushright}{\footnotesize \par}

\begin{flushright}
\rotatebox{90}{\footnotesize\quad Lateral view} \includegraphics[width=0.7\linewidth]{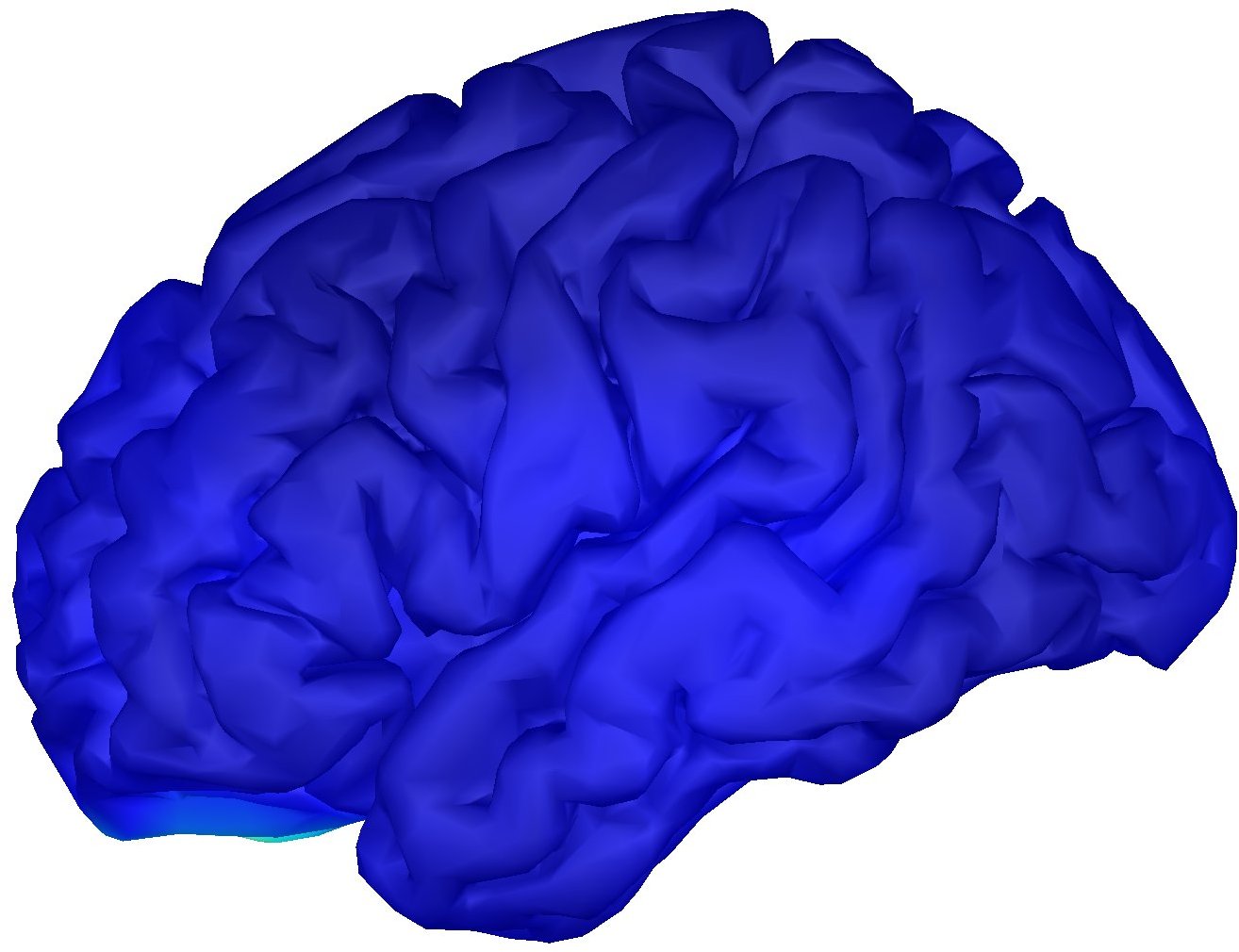}
\par\end{flushright}%
\end{minipage}%
\begin{minipage}[t]{0.49\linewidth}%
\begin{flushleft}
{\footnotesize{}\quad{}Right hemisphere}
\par\end{flushleft}{\footnotesize \par}

\begin{flushleft}
\includegraphics[width=0.7\linewidth]{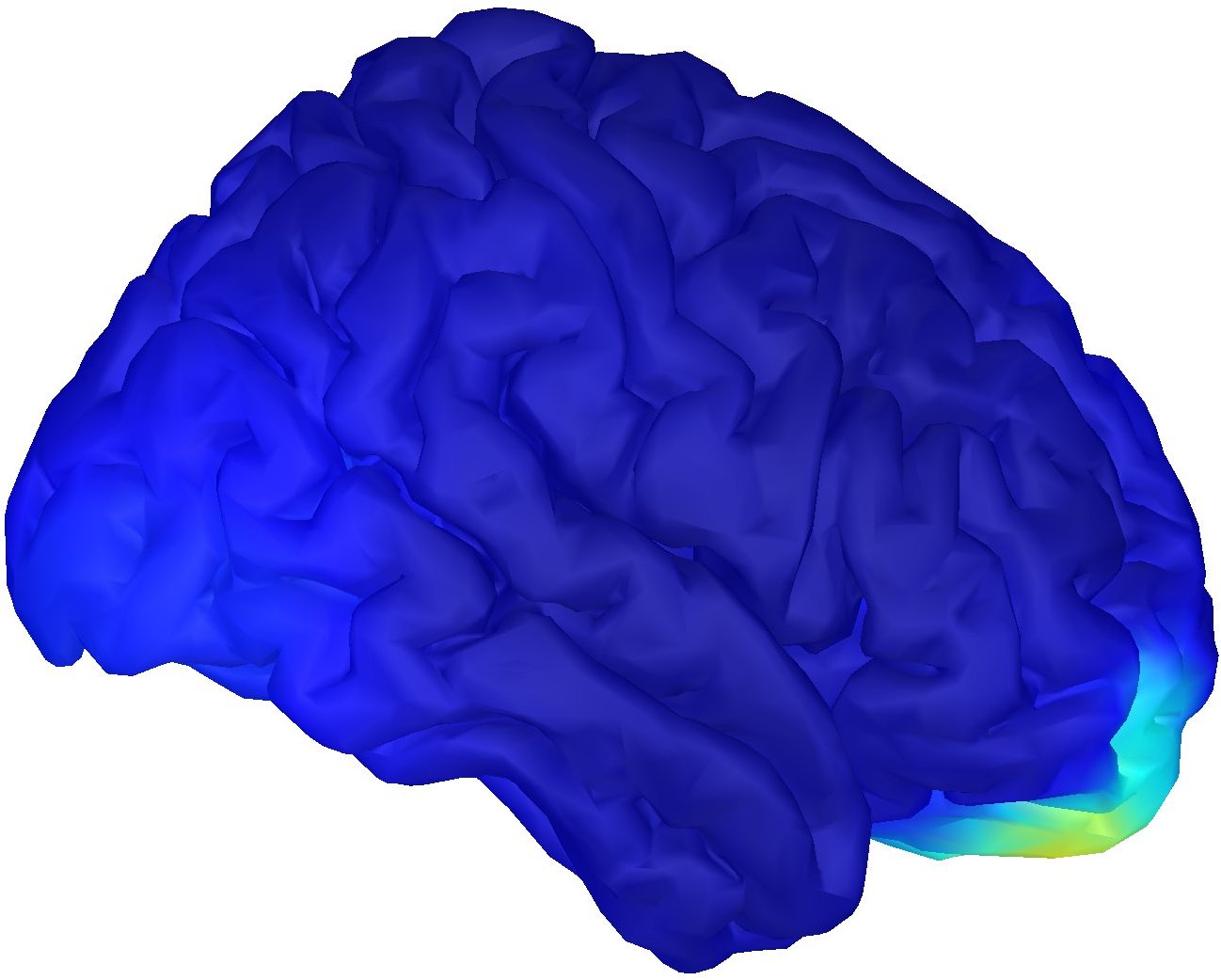}
\par\end{flushleft}%
\end{minipage}

\begin{minipage}[t]{0.49\linewidth}%
\begin{flushright}
\rotatebox{90}{\footnotesize\quad Medial view} \includegraphics[width=0.7\linewidth]{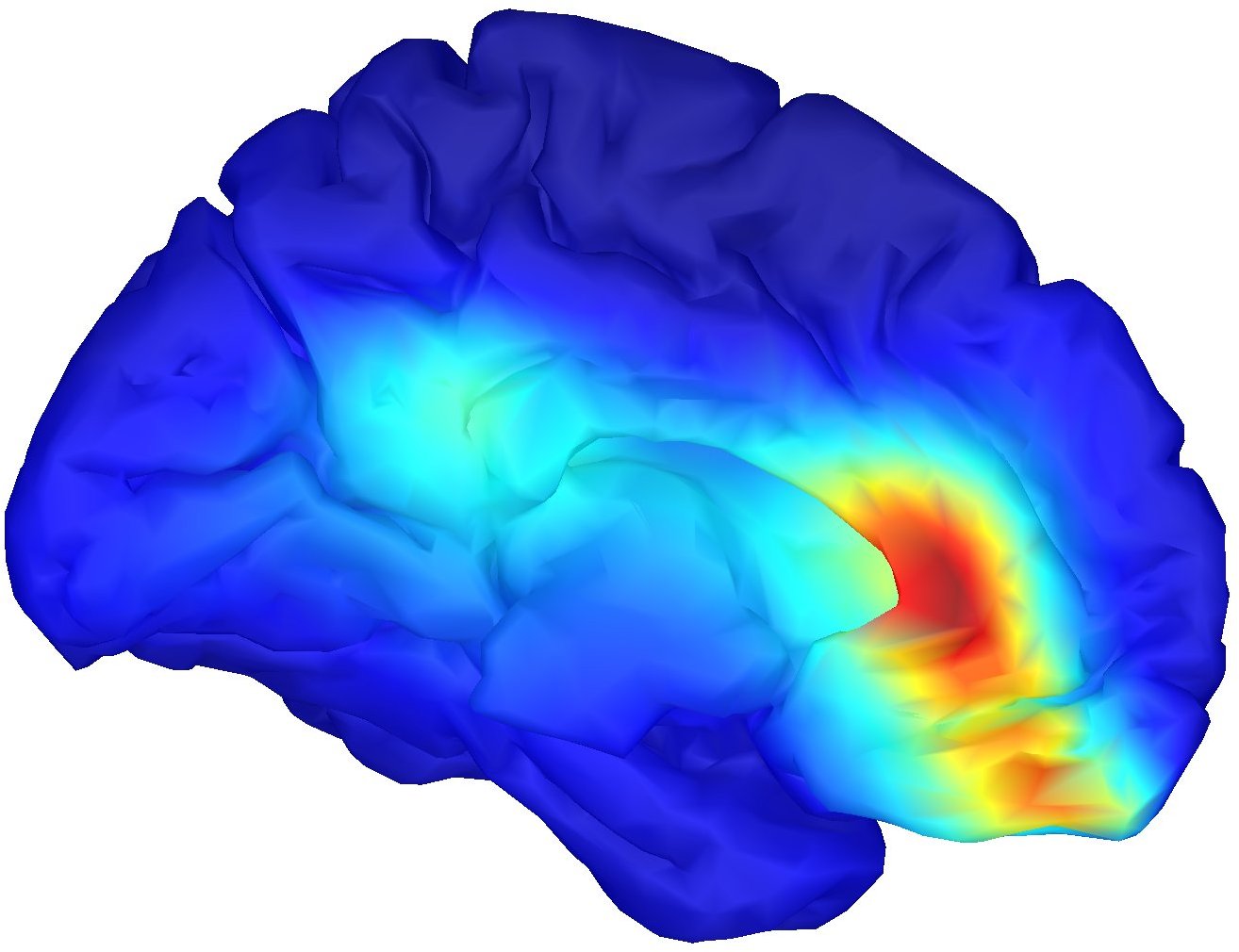}
\par\end{flushright}%
\end{minipage}%
\begin{minipage}[t]{0.49\linewidth}%
\begin{flushleft}
\includegraphics[width=0.7\linewidth]{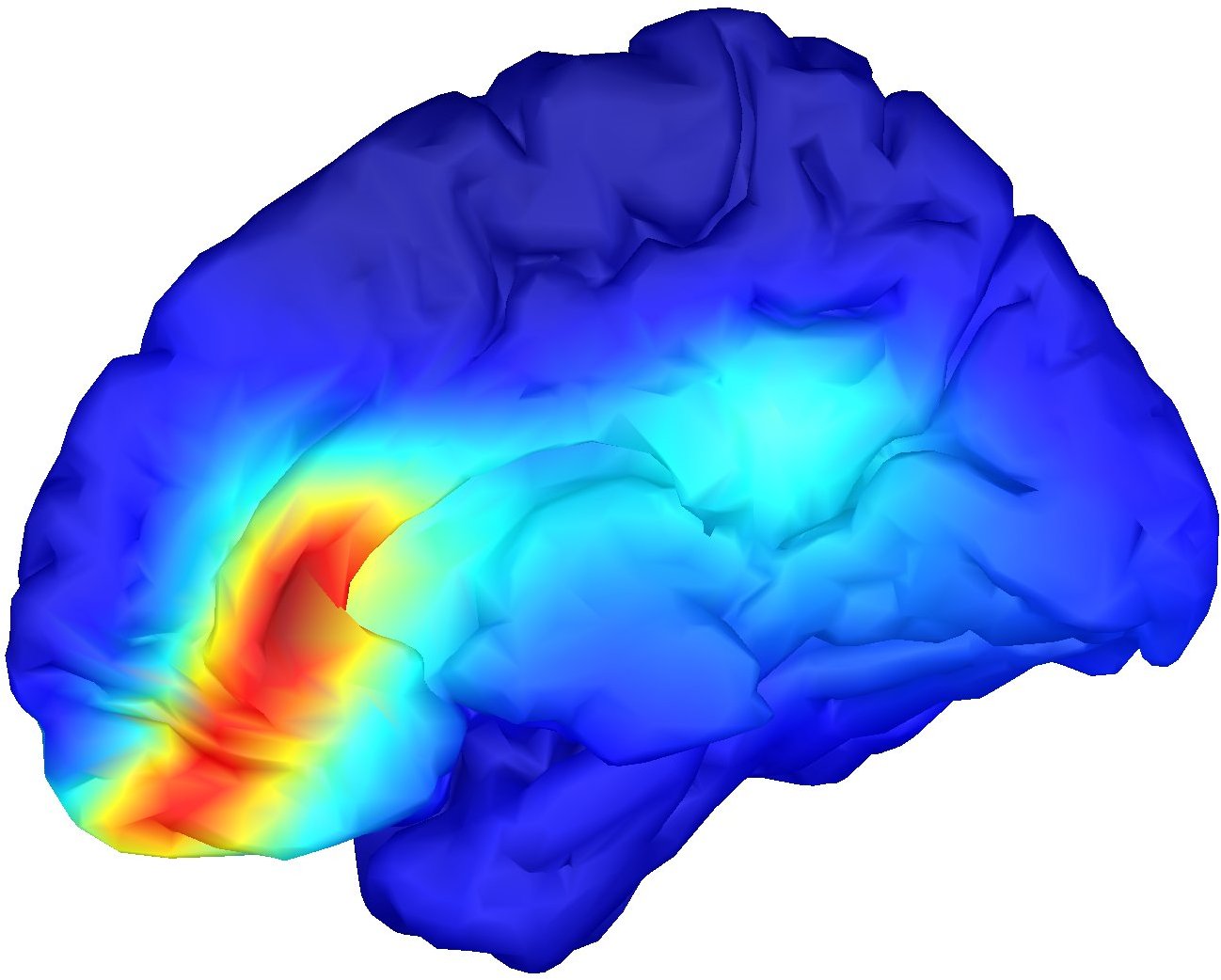}
\par\end{flushleft}%
\end{minipage}
\centerline{(b) linear}

\caption{Group averaged dipole involvement corresponding to the spatial filters
identified by the (a) non-linear and (b) linear MERLiN algorithm; lateral and medial views of the left and right hemisphere. (All colour scales from ``blue'' to ``red''
range from $0$ to the largest value to be plotted.)}

\label{fig:brain2}
\end{figure}

The group averaged results of our extended algorithm are depicted in Figure
\ref{fig:brain2}.(a). Similar to the results in \cite{grosse2015identification} and the results we obtained with
the linear MERLiN algorithm (cf. Figure \ref{fig:brain2}.(b)) the analysis indicates that the MPC is a causal effect of
the SPC. The non-linear method yields results that are
in high accordance with the ones obtained by our linear method while exhaustive search additionally revealed the anterior middle frontal gyrus as effect of the SPC.

\section{Conclusions}

We have developed the \emph{non-linear }MERLiN algorithm that is able
to recover a causal effect from an observed linear mixture with no
constraint on the functional form of this cause-effect relationship.
Iteratively projecting out directions and applying the MERLiN algorithm may allow to identify multiple distinct causal effects.
For EEG data we found no qualitative difference
to the linear method, which indicates that linear methods suffice to identify within-frequency cause-effect relationships in EEG data. Future research will focus on theoretical
analysis of the presented methods and assumptions and investigate
applicability to other real world data.

\bibliographystyle{IEEEtran}
\bibliography{references}

\end{document}